\documentclass[twocolumn,prl,showpacs,amsmath,amssymb]{revtex4}

\usepackage{graphics,graphpap,amssymb,psfrag,subfigure,picinpar}
\usepackage{graphicx}
\usepackage{hyperref}

\begin{document}

\title{The Effect of Quantum Phase Slip Interactions on the Transport of Thin Superconducting Wires}

\author{{\sc Dganit Meidan$^1$, Yuval Oreg$^1$, and Gil Refael$^2$}\\
{\small \em $^1$Department of Condensed Matter Physics, Weizmann Institute of Science, Rehovot, 76100, ISRAEL\\
$^2$ Department of Physics, California Institute of Technology,
Pasadena, California 91125, USA}}

\begin{abstract}

We study theoretically the effect of interactions between quantum
phase slips in a short superconducting wire beyond the dilute
phase slip approximation. In contrast to the smooth transition in
dissipative Josephson junctions, our analysis shows that treating
these interactions in a self consistent manner leads to a very
sharp transition with a critical resistance of $R_Q= h/(4e^2)$.
The addition of the quasi-particles resistance at finite
temperature leads to a quantitative agreement with recent
experiments on short MoGe nanowires. Our treatment is
complementary to the theory of the thermal activation of phase
slips, which is only valid for temperature at the vicinity of the
mean field metal to superconductor transition. This self
consistent treatment should also be applicable to other physical
systems that can be mapped onto similar sine-Gordon models, in the
intermediate-coupling regime.
\end{abstract}
\pacs{74.78.Na, 74.20.-z, 74.40.+k, 73.21.Hb}
 \maketitle

One of the most intriguing problems in low-dimensional
superconductivity is the understanding of the mechanism that
drives the superconductor-insulator transition (SIT). Experiments
conducted on quasi-one-dimensional (1D) systems have shown that
varying the resistances and diameters of thin metallic wires can
suppress superconductivity \cite{PXiongPRL1997,YOregPRL1999}, and
in certain cases lead to an insulating-like behavior
\cite{FSharifiPRL1993,ABezryadinNAT2000,CNLauPRL2001,ATBollinger2005}.
We focus on recent experiments conducted on short ultrathin MoGe
nanowires \cite{ATBollinger2005} that show a SIT driven by the
wires' normal state resistance with a critical resistance
$R_c\approx R_Q$.

The universal critical resistance may suggest that at a
temperature much lower than the mean field transition temperature,
$T\ll T_c $, the wire acts as a superconducting (SC) weak link
resembling a Josephson junction (JJ) connecting two
superconducting leads. Schmid~\cite{ASchmidPRL1983} and
Chakravarty \cite{SChakravartyPRL1982} showed that a JJ undergoes
a SIT as a function of the junction's shunt resistance,
$R_\textrm{s}$, with a critical resistance of $R_Q$, and that the
temperature dependence of the resistance obeys the power law
$R(T)\propto T^{2\left(\frac{R_Q}{R_\textrm{s}} -1\right)}$. The
theory was later extended to JJ arrays and SC wires
\cite{HPBuchlerPRL2004,GRefael2005,GRefael2006}. Within these
theories, a similar power-law temperature dependence of the
resistance at low temperature prevails.

A power-law $R(T)$ implies a rather wide SIT, with a resistance
that has only a weak temperature dependence in the critical
region. Contrary to the theoretical analogy with a single JJ,
A.~T.~Bollinger \textit{et al.} \cite{ATBollinger2005} observe
that the resistance of quasi 1D MoGe nanowires exhibits a strong
temperature dependence, even close to the SIT. In fact, the
resistance could be fitted with a modified LAMH theory
\cite{JSLangerPR1967,DEMcCumberPRB1970} of thermally activated
phase slips down to very low temperatures. Nevertheless, for these
narrow wires it appears that the LAMH theory is valid only in a
narrow temperature window
\cite{CommentLAMHValidity,RSNewbowerPRB1972}. Moreover, the LAMH
analysis does not explain the appearance of the universal critical
resistance $R_Q $.

In this manuscript we propose an approach that captures both the
universal critical resistance at the SIT, and the sharp decay of
the resistance as a function of temperature. As in previous works,
we treat the SIT in nanowires as a transition governed by quantum
phase-slip (QPS) proliferation. This picture alone, however,
cannot account for the observed strong temperature dependence of
the resistance. We claim that the key ingredient left out in
previous works is the inclusion of interactions between phase
slips in the finite-size wire, when the phase-slip population is
dense - a situation that occurs in thin wires at high temperatures
(see also Ref. \cite{GRefael2006}).

We treat these interactions in a mean-field type approximation:
when analyzing the behavior of a small segment of the wire, we
include in its effective shunt-resistance the resistance due to
phase slips elsewhere in the wire. This scheme is motivated by
numerical analysis of a related problem, an interacting pair of
resistively-shunted JJs \cite{PWernerJSM2005}. Primarily, this
self-consistent treatment produces the sharp temperature
dependence of the resistance. In addition, we include the effects
of the Bogoliubov quasi-particles, which couple to the potential
gradient created by each
phase-slip~\cite{CommentBdGQuasiParticles}. Consequently, the
resistance obtained in the experiment can be fitted without
resorting to the LAMH theory beyond its limit of validity. This
self-consistent approximation of phase-slip interactions should be
applicable to similar multiple sine-Gordon models in the
intermediate-coupling regime.

The microscopic action for a SC wire can be obtained from the BCS
Hamiltonian by a Hubbard-Stratanovich transformation followed by
an expansion around the saddle point \cite{CommentDirtyLimit}. In
the limit of low energy scales, $\omega, Dq^2~\ll~\Delta_0$, this
yields \cite{DSGolubevPRB2001}:
\begin{eqnarray}\label{effective_action}
\nonumber
  S &=& N_0A\Delta_0^2\int_0^L dx\int_0^{1/T} d\tau\left\{\frac{\rho^2}{2}\left[\ln\left(\rho^2\right)-1\right] \right.\\
  &&\left. +2\xi_0^2\rho^2\left[\phi'^2+\frac{1}{v_\phi^2}\dot{\phi}^2 \right]+\xi_0^2\left[\rho'^2+\frac{1}{v_{\rho}^2}\dot{\rho}^2
  \right]\right\},
\end{eqnarray}
where $L$ and $A$ are the wire's length and cross section,
respectively, $\xi_0^2=\pi D/8\Delta_0 $,
$v_\rho=\sqrt{(3\pi/2)D\Delta_0}$ the amplitude velocity,
$v_\phi=\sqrt{\pi D\Delta_0(2AV_cN_0+1)} $ the phase velocity,
$V_c$ the Fourier transform of the short range Coulomb
interaction, $N_0$ the density of states, $D$ the electronic
diffusive constant in the normal state, and the SC order parameter
is parameterized as $\Delta = \Delta_0\rho e^{i\phi}$, with
$\Delta_0$, the mean field solution. For the wires in
Ref.~\onlinecite{ATBollinger2005} $2AV_cN_0\propto N_\bot\gg 1 $,
leading to $v_\rho\ll v_\phi\propto v_\rho\sqrt{N_\bot}$. Here
$N_\bot=p_F^2A /\pi^2$ is the number of 1D channels in the wire.

This action supports QPS excitations, which are characterized by
two distinct length scales: $v_\rho/\Delta_0\propto \xi\ll
\xi\sqrt{N_\bot}\propto v_\phi/\Delta_0$. For very long wires, $
\xi\ll \xi \sqrt{N_\bot}\ll L$, in the dilute phase slip
approximation, this problem can be mapped onto the perturbative
limit of the $1+1$ sine-Gordon model \cite{HPBuchlerPRL2004}. In
the opposite limit of very short wires, $ L<\xi\ll \xi
\sqrt{N_\bot}$, the system resembles a JJ and can be mapped onto
the $0+1$ sine-Gordon model. However the MoGe nanowires in Ref.
\onlinecite{ATBollinger2005} appear to be in the intermediate
regime, $\xi \ll L\ll \xi \sqrt{N_\bot}$. Hence, while phase slips
occur in different sections of the wire, they are
indistinguishable, as each creates a phase fluctuation that
spreads over distances larger than the wire itself.

Moreover, the wires in Ref. \cite{ATBollinger2005} have a sizable
bare fugacity. Using Eq. (\ref{effective_action}), one can
estimate the core action of a phase slip of duration $\tau_0 =
1/\Delta_0$ by minimizing the action with respect to the diameter
of a phase slip \cite{CommentCoreAction,RSmithUnPublished}. This
yields a core action of $S_c
=\pi/8\sqrt{1/3(1-\epsilon)}R_Q/R_\xi$. Here $\epsilon = 1-T/T_c$
is the reduced temperature and $R_\xi= R_{\textrm{W}} \xi/L $ is
the normal resistance of a section of length $\xi$. With this
expression, the measured values of $R_{\textrm{W}}$ and $L$ (Table
\ref{param_table}), and the assumption that $\xi\approx 20 nm$,
the phase slip fugacity is estimated by $\zeta = e^{-S_c}\approx
0.05-0.45$, for the different wires. Namely, in the critical
region, $R_{\textrm{W}} \approx R_Q $, there is a dense population
of phase slips which interact with one another; thus the dilute
phase slip approximation is no longer a proper description.

The flow equation for the fugacity of a phase slip anywhere in the
wire is given by:
\begin{eqnarray}\label{RG_fugacity}
  d\zeta/dl &=& \left(1-
   R_Q/R_\textrm{s}\right)\zeta,
\end{eqnarray}
where $dl = -d\ln\Lambda $, and $\Lambda $ is the running RG
scale. Eq. (\ref{RG_fugacity}) treats the phase slip as occuring
on an effective JJ, with $R_\textrm{s}$ being the effective
shunting resistance of the entire wire. When $\zeta$ is small,
$R_\textrm{s}$ will include only the effective impedance of the
leads. If $\zeta$ is not very small, we will need to include in
Eq.~(\ref{RG_fugacity}) additional terms of higher powers of
$\zeta$, which describe interactions between QPSs. Currently,
there is no complete understanding of how to approach the field
theory of phase slips in the regime of intermediate fugacity
$\zeta\lesssim 1$. Therefore, we suggest to take into account a
finite $ \zeta$ by including the resistance due to other phase
slips in the wire in the effective shunt resistance of the
junction, $R_\textrm{s}$. This treatment was successfully tested
by numerical analysis of a simpler analog of the system, an
interacting pair of resistively shunted JJs \cite{PWernerJSM2005}.
Indeed, including the interaction between phase slips as a
modification of the shunt, $R_{\textrm{s}}$, is akin to guessing
the form of a complete resummation of higher-order $\zeta$ terms
in
Eq.~(\ref{RG_fugacity})~\cite{CommentNextOrderInRG,BulgadaevPLA1981}.

The main physical intricacy is the determination of the effective
shunt resistance, $R_\textrm{s}(\zeta)$, that governs the
renormalization of the phase slip fugacity [Eq.
(\ref{RG_fugacity})]. A phase slip produces time-varying phase
gradients, and hence electrical fields. These dissipate through
two channels in parallel: the SC channel - which has an effective
resistance due to other phase slips, $R_\textrm{{ps}}$ - and the
quasi-particles conducting channel
\cite{CommentBdGQuasiParticles}, $R_\textrm{{qp}} $. Once the
disturbance reaches the leads, it also dissipates through the
electro-dynamical modes of the large electrodes, whose real
impedance is parameterized by $R_\textrm{{elec}}$.

\begin{figure}[h]
\begin{center}
\includegraphics[width=0.45\textwidth]{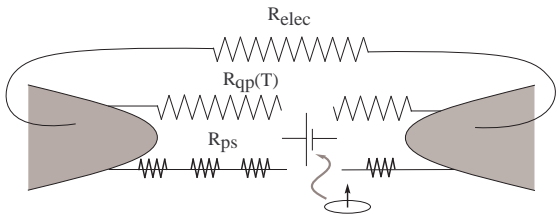}
\caption[0.6\textwidth]{ The effective electrical circuit for the
nanowire. A phase slip produces time-varying phase gradients, and
hence electrical fields, that dissipate through the
quasi-particles conducting channel, $R_\textrm{{qp}} $, and
through the SC channel, that has an effective resistance due to
other phase slips, $R_\textrm{{ps}} $. Once the disturbance
reaches the edges, it also dissipates through the
electro-dynamical modes of the large electrodes, whose real
impedance is represented by $R_\textrm{{elec}} $. }
\label{fig:circuit}
\end{center}
\end{figure}

For $T\ll~T_c $, the resistance of the quasi particles,
$R_\textrm{{qp}} $, can be approximated by $R_\textrm{{qp}} =
\frac{m}{e^2\tau_{\textrm{n}} n_\textrm{{qp}}}\frac{L}{A}=
R_{\textrm{n}} \frac{n}{n_\textrm{{qp}}}
   \approx R_{\textrm{n}} \sqrt{\frac{T}{2\pi
   \Delta_0}}e^{\frac{\Delta_0}{T}}
 $.
Unfortunately, we lack a microscopic model for the impedance of
the electrodes, as this depends on the details of the system.
However, we expect that at large scales, $T<\Lambda<\Delta $, the
electrodes will act as a transmission line to the electromagnetic
waves generated by the phase slip. This transmission line is
characterized by a real impedance which we denote as
$R_\textrm{{elec}}$, and use as a fitting parameter. Hence, the
effective shunt resistance that affects the renormalization of the
phase slip fugacity at $T<\Lambda<\Delta $ [Eq.
(\ref{RG_fugacity})] is
\begin{equation}\label{effective_shunt_resistance}
    R_\textrm{s}[\zeta(\Lambda)] =
    R_\textrm{{elec}}+1/\left(1/R_\textrm{{ps}}[\zeta(\Lambda)]+1/R_\textrm{{qp}}(T)\right).
\end{equation}Fig. \ref{fig:circuit} shows the circuit we suggest describes the
system.

The resistance is measured in response to an applied DC current.
At this zero frequency limit, the electrodes act as a capacitor
connected in parallel to the wire. Therefore, the measured
resistance is the total wire resistance, unaffected by the
environment, which is cut off from the wire.
\begin{equation}\label{total_DC_resistance}
  R_\textrm{{tot}}(T)= 1/\left(1/R_\textrm{{ps}}[\zeta(T)]+1/R_\textrm{{qp}}(T)\right).
\end{equation}

The occurrence of a phase slip causes a resistance in the
otherwise SC wire through the relation $R_\textrm{{ps}}\propto
(L/\xi) \zeta^2 $. Using this relation and Eq.
(\ref{RG_fugacity}), we can write an RG equation for the
dimensionless resistance
 \begin{eqnarray}\label{RG_resistance}
 \nonumber
   \frac{d\zeta^2}{dl} &=& 2\zeta\frac{d\zeta}{dl} =2 \left(1- \frac{R_Q}{R_\textrm{{s}}(\zeta)}\right)\zeta^2 \\
   \Rightarrow\frac{d\left(R_\textrm{{ps}}/R_Q\right)}{dl} &=& 2 \left(1-
   \frac{R_Q}{R_\textrm{{s}}(R_\textrm{{ps}})}\right)\left(\frac{R_\textrm{{ps}}}{R_Q}\right).
 \end{eqnarray}
If we now substitute the effective resistance in Eq.
(\ref{effective_shunt_resistance}), depicted in Fig.
\ref{fig:circuit}, into the RG equation (\ref{RG_resistance}), we
find the following self consistent equation:
\begin{eqnarray}\label{self_consistancy}
   \frac{d\left(\frac{R_\textrm{{ps}}}{R_Q}\right)}{dl} &=&\!\! 2\!\!\left[1-\frac{R_Q}
   {\left(R_\textrm{{elec}}+\frac{R_\textrm{{ps}}
   R_\textrm{{qp}}(T)}{R_\textrm{{ps}}+R_\textrm{{qp}}(T)}\right)}\right]\!\!\left(\frac{R_\textrm{{ps}}}{R_Q}\right).
 \end{eqnarray}
Integrating Eq.(\ref{self_consistancy}) from the ultra violet (UV)
cutoff, $\Delta(T^*)=T^*$, to the infra red cutoff, $T$,
yields~$R_\textrm{{ps}}(T) $.

\begin{figure}[h]
\begin{center}
\includegraphics[width=0.4\textwidth]{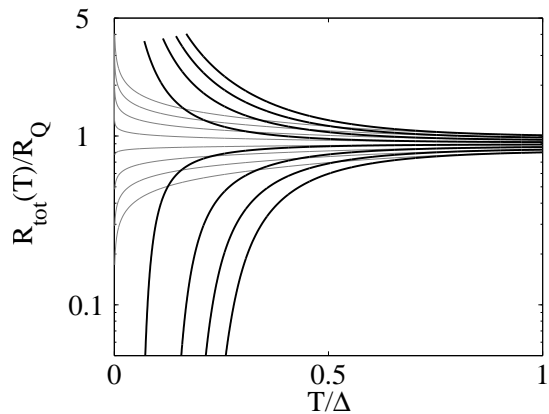}
\caption[0.6\textwidth]{Reduced resistance $R_\textrm{{tot}}/R_Q $
as a function of renormalized temperature $T/\Delta $ for
different $R^*/R_Q$, ranging from $0.8$ (lower plot) to $1.01$
(upper plot) in increasing steps of $0.03$. The environment
resistance was taken to be $R_\textrm{{elec}}/R_Q= 0.1$. Gray
traces indicate the total wire resistance
$R_\textrm{{tot}}=\frac{R_\textrm{{ps}}
R_\textrm{{qp}}(T)}{R_\textrm{{ps}}+R_\textrm{{qp}}(T)} $, with
$R_\textrm{{ps}}(T) =
R^*\left(T/\Delta\right)^{2(R_Q/R_\textrm{s}-1)}$, and constant
$R_s=R^*+ R_\textrm{{elec}}$. The insulating plots are cutoff at
temperature, $T_0 $, for which $\zeta(T_0) = 1 $. }
\label{fig:Rtot_self_consistent_vs_non}
\end{center}
\end{figure}

The wire's DC resistance Eq. (\ref{total_DC_resistance}),
calculated using Eq. (\ref{self_consistancy}), is plotted in
Fig~\ref{fig:Rtot_self_consistent_vs_non} as a function of
temperature for different $R^*$. Here $R^*\equiv
R_\textrm{{tot}}(T^*)$ is the normal state resistance of the wire,
at the UV cutoff $\Delta(T^*)=T^*$. We assume that the wires are
thin enough such that the mean field transition from normal to SC
is wide, and $R^*\approx R_\textrm{{tot}}(T_c^+)=R_{\textrm{n}} $.
For simplicity, we have assumed throughout our calculation that
$\Delta(T)\approx\Delta $. This assumption holds at the low
temperature regime, $T\ll \Delta$, where the theory of QPSs, based
on the effective action Eq. (\ref{effective_action}), is expected
to be valid. The resistance of the environment is taken to be
$R_\textrm{{elec}}=0.1 R_Q$. In practice, $\Delta$, $R^*$ and
$R_\textrm{{elec}}$ can be used as fitting parameters, when
comparing the theory to the experiment. Moreover, $R^* $ can be
determined independently as the resistance measured below the drop
that indicates passing through $T_c$ of the SC films (see Ref.
\onlinecite{ATBollinger2005}).

Eq. (\ref{self_consistancy}) is also applicable to a wire with a
normal resistance larger than the quantum resistance, $R^*>R_Q$.
In this limit the fugacity increases in the renormalization
process, and Eq. (\ref{RG_fugacity}) is no longer valid for
$\zeta\gtrsim 1 $. For wires with $R^*>R_Q $, we overestimate the
phase slip fugacity as $\zeta(T^*)=0.5 $, and plot the reduced
resistance, $R_\textrm{{tot}}/R_Q $, down to the temperature,
$T_0$, for which $\zeta(T_0)=1$.  The results are shown in
Fig.~\ref{fig:Rtot_self_consistent_vs_non}. Overestimating
$\zeta(T^*)$ gives an upper bound on $T_0$, where Eq.
(\ref{RG_fugacity}) is no longer applicable. Fig.
\ref{fig:Rtot_self_consistent_vs_non} shows that the transition
between SC and insulating wires occurs for a critical resistance
$R_c\approx R_Q$. But in contrast to the standard Josephson
junction theory, (gray curves in Fig.
\ref{fig:Rtot_self_consistent_vs_non}), the transition is much
more sharp (notice the $\log $ scale).

\begin{table}[ht]
\begin{center}
\begin{tabular}{| c | r  r @{.} l | r @{.} l  r @{.} l  r @{.} l |}
\hline
 &\multicolumn{3}{|c|}{Measured
Values}&\multicolumn{6}{|c|}{Fitting Values}\\
\hline
Curve & $L (nm)$ &\multicolumn{2}{c|}{$R_{\textrm{W}} (k\Omega)$} &\multicolumn{2}{c}{$R^* (k\Omega)$} &\multicolumn{2}{c}{$\Delta(K)$} &\multicolumn{2}{c|}{$R_\textrm{{elec}} (k\Omega)$}  \\
\hline
$1$ & 177\phantom{0} & \phantom{00}5&46 & \phantom{00}4&2 &  \phantom{0}2&5 & \phantom{00}1&2\\
$2$ & 43\phantom{0} & 3&62 & 2&62 & 2&35 & 1&25\\
$3$ & 63\phantom{0} & 2&78 & 2&13 & 3&07 & 0&66\\
$4$ & 93\phantom{0} & 3&59 & 2&89 & 3&85 & 0&55\\
$5$ & 187\phantom{0} & 4&29 & 4&5 & 6&55 & 0&31\\
$6$ & 99\phantom{0} & 2&39 & 2&09 & 4&84 & 0&4\\
\hline
\end{tabular}
\end{center}
\caption{Summary of nanowire parameters, and the parameters used
to fit the experimental data; $L$ is the length of the wire
determined from SEM images, $ R_{\textrm{W}}$ is the wire's normal
state resistance, taken as the resistance measured below the film
transition. Fitting parameters: $R^*=R_{\textrm{tot}}(T = \Delta)
$ is the wire's resistance at the UV cutoff, $\Delta $ the SC
order parameter, and $ R_\textrm{{elec}}$ the impedance of the
electrodes at $T<\Lambda<\Delta $.} \label{param_table}
\end{table}

\begin{figure}[h]
\begin{center}
\includegraphics[width=0.5\textwidth]{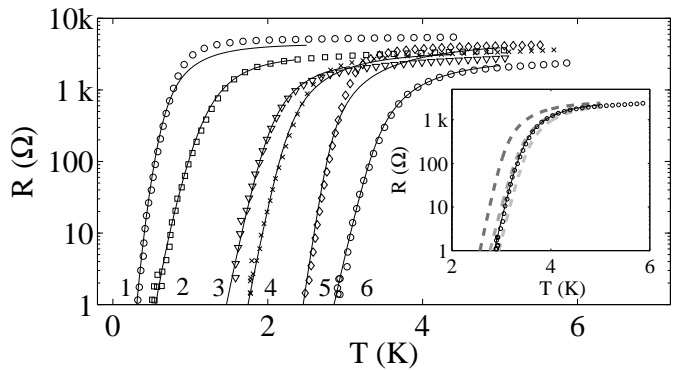}
\caption[0.6\textwidth]{Comparison between the theory (solid line)
and the experimental data \cite{ATBollinger2005}. Details of the
wires, as well as fitting parameters are summarized in Table
\ref{param_table}. Note that the estimated values for $R^*$ are in
good agreement with the measured values of the resistance of the
wire at $T=\Delta $. The discrepancies between theory and
experiment occur for the longer wires (curves $1 $ and $ 5$). In
these wires the condition $v_\phi/\Delta_0 \gg L$ might not hold,
and one should consider additional renormalization of the fugacity
at higher scales. Inset: The effect of an adjustment of the fit
parameters, in roughly $10\%$ (from dark gray to light gray):
$\Delta = 4.356 K$, $R_\textrm{{elec}}=440 \Omega$ and $R^* = 1881
\Omega $.} \label{fig:fit}
\end{center}
\end{figure}

A comparison between the theoretical curves and the experimental
data taken from Ref. \onlinecite{ATBollinger2005} is shown in
Fig.~ \ref{fig:fit}. The curves were calculated by fitting $R^* $,
$\Delta$, and $R_\textrm{{elec}}$. Since the theory of QPSs is
expected to be valid at $T\ll \Delta $, deviations from the
theoretical curves at high temperature are reasonable. In general,
as $\Delta$ is proportional to $T_c $, increasing $\Delta$ shifts
the sharp decay of the resistance to high temperatures. Both $R^*$
and $R_\textrm{{elec}}$ affect the high temperature resistance,
whereas $R^*$ and $\Delta $ control the width of the transition.

We have made an attempt to fit the experimental data corresponding
to the insulating wires of Ref. \onlinecite{ATBollinger2005}. As
the insulating wires are thinner, we expect a strong suppression
of $T_c$ \cite{YOregPRL1999}. Consequently, the theory of QPSs
with $\zeta\lesssim 1$, is valid in a narrow range of
temperatures. While we manage to fit the experimental curves in
this regime of parameters, we do not find it of scientific merit
to present the results (our fits cover a range of $\sim 10 $ data
points, with three fitting parameters).

\vspace{-.3cm} In conclusion, we have studied theoretically the
effect of interactions between QPSs, in short SC wires, beyond the
dilute phase slip approximation. Our analysis shows that treating
these interactions in a self consistent manner, produces a sharp
SIT transition, with a critical resistance $R_c\approx R_Q$, in
agreement with recent experiments conducted on short ultrathin
MoGe nanowires \cite{ATBollinger2005}. Moreover, we have shown
that adding the resistance due to the presence of a finite number
of BdG quasi-particles in the wire leads to a quantitative
agreement with the experimental curves. Our method should be
applicable to a wider range of physical problems which involve the
proliferation of topological defects with a sizable bare fugacity.
In particular, it could be applied to the study of a Luttinger
liquid with an extended impurity \cite{CLKanePRB1992}.

We would like to thank E.~Demler, R.~A.~Smith, and P. Werner.
Special thanks to A.~Bezryadin for making his data available to
us. This study was supported by a DIP grant. \vspace{-.5cm}

\bibliographystyle {h-physrev3}

\begin{thebibliography}{10}
\bibitem{CommentBdGQuasiParticles}
This could be seen by considering a time dependent Bogoliubov de
Gennes (BdG)
  equation, with a time dependent phase: $\Delta =
  |\Delta|\exp\left(i\phi(t)\right)$. By the canonical transformation $(u',
  v')=\exp\left(\frac{i}{2}\sigma_z\phi\right)(u, v) $, we produce an effective
  potential $\dot{\phi}/2 $ for the quasi-particles in the BdG Hamiltonian.
\bibitem{CommentDirtyLimit}
The expansion is done in the dirty limit, $\Delta_0\tau\ll 1$,
where $\tau$ is
  the scattering mean free time.
\bibitem{CommentCoreAction}
To estimate the core action, we choose the following trial
function for a phase
  slip: $\phi = \arctan{\left(v_{\phi}\tau/x\right)} $, and
  $\rho=\min{[\sqrt{\left(x/x_0\right)^2+ \left(\Delta_0\tau\right)^2},1]} $.
  Identifying the part of the action [Eq. (\ref{effective_action})], that
  corresponds to $\rho \neq 1$ as the core action, we minimize this expression
  with respect to the phase slip diameter, $x_0 $. Keeping only leading terms
  in $N_\bot=\left(p_F^2A /\pi^2\right)$, one finds $Sc
  =\pi/8\sqrt{1/3(1-\epsilon)}R_Q/R_\xi$.
\end{thebibliography}

\newcommand{\noopsort}[1]{} \newcommand{\printfirst}[2]{#1}
\newcommand{\singleletter}[1]{#1} \newcommand{\switchargs}[2]{#2#1}
\providecommand{\bysame}{\leavevmode\hbox
to3em{\hrulefill}\thinspace}
\providecommand{\MR}{\relax\ifhmode\unskip\space\fi MR }
\providecommand{\MRhref}[2]{%
  \href{http://www.ams.org/mathscinet-getitem?mr=#1}{#2}
} \providecommand{\href}[2]{#2}

\vspace{-.2cm}
\end{document}